\documentclass[aip,jcp,reprint]{revtex4-1}

\usepackage{graphicx}
\usepackage{amsmath,amssymb,amsthm}

\renewcommand{\vec}[1]{\boldsymbol{#1}}
\newcommand{\beq}{\begin{equation}}
\newcommand{\eeq}{\end{equation}}
\newcommand{\beqa}{\begin{eqnarray}}
\newcommand{\eeqa}{\end{eqnarray}}
\newcommand{\e}{\mathrm{e}}
\newcommand{\w}{\omega}

\newcommand{\ket}[1]{\left| #1 \right\rangle}

\newcommand{\der}[2]{\frac{\mathrm{d}#1}{\mathrm{d}#2}}
\newcommand{\ketbra}[2]{\left|#1\right\rangle\hskip-1mm\left\langle #2\right|}

\newcommand{\ketbrai}[3]{\left|#1\right\rangle_{#3}\hskip-1mm\left\langle #2\right|}

\begin{document}

\title{Consistent treatment of coherent and incoherent energy transfer dynamics using a variational master equation} 
\author{Dara P. S. McCutcheon} \email{dara.mccutcheon@ucl.ac.uk}
\affiliation{Department of Physics and Astronomy, University College London, Gower Street, London WC1E 6BT, UK}
\affiliation{London Centre for Nanotechnology, University College London}
\author{Ahsan Nazir} \email{a.nazir@imperial.ac.uk}
\affiliation{Blackett Laboratory, Imperial College London, London SW7 2AZ, UK}
\affiliation{Department of Physics and Astronomy, University College London, Gower Street, London WC1E 6BT, UK}

\date{\today}

\begin{abstract}

We investigate the energy transfer dynamics in a donor-acceptor model by developing 
a time-local master equation technique based on a variational transformation of the underlying Hamiltonian. 
The variational transformation allows a minimisation of the Hamiltonian perturbation term 
dependent on the system parameters, and consequently results in a versatile master equation valid over 
a range of system-bath coupling strengths, temperatures, and environmental spectral densities. While our formalism reduces to the well-known Redfield, 
F\"{o}rster and polaron forms in the appropriate limits, in general it is not equivalent to perturbing in either the system-environment or donor-acceptor coupling 
strengths, and hence can provide reliable results between these limits as well. Moreover, we show how to include the effects of both environmental correlations and non-equilibrium 
preparations within the formalism.

\end{abstract}

\maketitle

\section{Introduction}
\label{intro}

The transfer of an electronic excitation 
is a ubiquitous process in physics, chemistry, and biology. Typically, 
coupling to the radiation field creates an excitation in one location (the donor), and through the exchange of a virtual photon, the 
excitation is passed to another (the acceptor).~\cite{andrews89} One of the many challenges in theoretically modelling such a process lies in correctly accounting for 
the influence of the environment surrounding the two (or more) sites, which determines whether the transfer is of a coherent or an incoherent nature. A common 
approach has been to assume that the energy transfer coupling strength is 
weak, and thus perform perturbation theory in this parameter. The resulting F\"{o}rster-Dexter rate equations then describe purely \emph{incoherent} energy transfer,~\cite{foerster59, dexter52} and are applicable in 
a wide variety of situations.~\cite{scholes03,olayacastro11}
However, recent experimental results providing evidence for \emph{quantum coherent} 
transfer in a number of systems~\cite{lee07,engel07, calhoun09, collini09sc,collini09,collini10,panitchayangkoona10,mercer09,womick09} have 
necessitated the development of theoretical techniques beyond this approximation. While extending F\"{o}rster theory to account for coherence within multi-site donors or acceptors is possible,~\cite{sumi99,scholes00,jang04} describing the \emph{dynamical} evolution of coherences between the donor and acceptor generally requires an alternative starting point.   

One such approach is to treat the system-environment coupling perturbatively, 
while accounting for the electronic coupling between the sites to all orders. Working with the resulting Redfield or Lindblad equations is vastly simpler than simulating the full dynamics,~\cite{b+p} and
has thus recently been put to great effect to analyse the interplay of dissipation and quantum coherence in the 
energy transfer dynamics of complex, many-site 
systems.~\cite{renger98,olayacastro08,mohseni08,plenio08,cao09,caruso09,rebentrost09,rebentrost09b,chin10,fassioli10,fassioli10_2,abramavicius10} However, by their very nature, treatments based upon a weak system-environment coupling assumption are often inadequate to describe systems strongly coupled to their environments and/or at high temperatures.~\cite{yang02,ishizaki09,ishizaki10,nalbach10_2} In this situation, there 
exists a number of (non-perturbative) numerical approaches which allow 
one to explore the full range of system-bath coupling and electronic transfer strengths, subject to certain method-specific constraints. Examples include density matrix renormalisation group,~\cite{prior10} 
numerical renormalisation group,~\cite{tornow08} and path integral~\cite{thorwart09,nalbach10} methods. 
Additionally, for specific forms (but not strengths) of system-environment coupling, numerically 
exact results can be obtained through the hierarchical equations of motion technique,~\cite{ishizaki09b,ishizaki09c,tanaka09,tanaka10,ishizaki10} which has recently been shown to be consistent with the path integral formalism.~\cite{nalbach11}

These methods have clear advantages in that their range of applicability is generally less restricted than those of the aforementioned perturbative approaches. However, given the 
physical insight perturbative techniques can impart, and the relative simplicity and efficiency with which they can be implemented, it remains of particular importance to develop approximation methods which allow one to probe as wide a range of parameter regimes as possible. For example, aspects of energy transfer dynamics have recently been explored using a time-local polaron-transform master equation technique,~\cite{nazir09,jang08,jang09,yang10,mccutcheon11,kolli11} 
which corresponds to perturbation theory in an environment-dressed electronic coupling term.~\cite{soules71,rackovsky73,abram75}  
For particular forms of the system-environment coupling, the polaron master equation %While it 
reduces to Redfield theory in the weak-coupling 
limit, %the polaron master equation 
but also remains valid for far greater temperatures and 
system-bath coupling strengths,~\cite{wurger98,jang09,mccutcheon10_2} and can thus also capture the F\"{o}rster limit. 
This allows, in particular, for a consistent exploration of the crossover from coherent to incoherent 
dynamics in energy transfer processes.~\cite{nazir09,mccutcheon11} The polaron method, however, 
suffers itself from a restriction to relatively small electronic coupling strengths 
(compared to typical frequencies in the environment), 
and, as mentioned, can only interpolate between the Redfield and F\"{o}rster limits for certain forms of system-environment coupling.

In this work we go beyond the polaron approach 
and present a master equation technique which combines a variationally-optimised unitary 
transformation~\cite{yarkony76,yarkony77,silbey84,harris85,silbey89} with the time-convolutionless projection operator formalism.~\cite{b+p} Modelling the environment as a bath of harmonic oscillators, the 
variational transformation displaces each 
mode according to the position of the excitation, and by an amount defined to minimise the resulting 
free energy of the interaction terms between the donor-acceptor pair and the environment. 
We contrast the resulting time-local master equation with the Redfield and polaron forms, and show that the variational approach is superior to both. Specifically, we show that while the variational master equation reproduces both the Redfield and polaron equations in the appropriate limits, it can also give qualitatively reliable results outside these parameter regimes. The variational master equation can therefore be used to 
explore energy transfer dynamics for a range of system-environment coupling forms, strengths, and environmental temperatures. 

The paper is organised as follows: In Section~\ref{Model} we introduce the donor-acceptor model and the variational transformation. We then outline the derivation of the time-local master equation in Section~\ref{MasterEquation}. In Section~\ref{dynamics} we use this master equation to explore the influence of both super-Ohmic and Ohmic environments on the donor-acceptor energy transfer dynamics. We also consider here the role of bath relaxation effects, and how they influence the rate of energy transfer. A brief summary is presented in Section~\ref{summary}, while Appendix~\ref{MEDerivation} gives further details of the master equation derivation, and Appendix~\ref{CorrelatedEnvironments} extends the formalism to consider correlated environmental fluctuations.~\cite{nazir09,mccutcheon11,nalbach10,fassioli10,hennebicq09,yu08,chen10,mccutcheon10,sarovar11,west10} 

\section{Model and Variational Transformation}
\label{Model}

We consider a donor-acceptor pair ($j=1,2$) each site of which we model as a two-level system with ground state 
$\ket{G}_j$ and excited state $\ket{X}_j$, split by an energy $\hbar\epsilon_j$. The pair interact via an electronic coupling of strength $V$ which transfers an excitation from one site to the other. To model the 
dephasing and dissipative effects of the environment, we linearly couple each excited state to a bath of 
harmonic oscillators.~\cite{ishizaki10,leggett87,weissbook} Although not the primary focus 
of this work, the formalism to be presented below can also been extended to include 
the effects of correlated fluctuations between the sites.~\cite{nazir09,mccutcheon11} This extension is outlined in Appendix~{\ref{CorrelatedEnvironments}}. 
In the absence of correlations, the total Hamiltonian reads (we set $\hbar=1$)
\begin{align}
H=&\sum_{j=1,2}\epsilon_j \ketbrai{X}{X}{j}+V\big(\ketbra{XG}{GX}+\ketbra{GX}{XG}\big)\nonumber\\
+&\sum_{j=1,2}\ketbrai{X}{X}{j}\sum_k g_{k , j}(b_{k, j}^{\dagger}+b_{k, j})+H_{B},
\label{TotalHamiltonian}
\end{align}
where $\ketbrai{X}{X}{1}=\ketbra{X}{X}\otimes(\ketbra{X}{X}+\ketbra{G}{G})$ and 
$\ketbrai{X}{X}{2}=(\ketbra{X}{X}+\ketbra{G}{G})\otimes\ketbra{X}{X}$ determine the excited state populations of the 
two sites, $g_{k,j}$ is the coupling constant of site $j$ to mode $k$ of bath $j$, described by creation (annihilation) operators 
$b_{k,j}^{\dagger}$ ($b_{k,j}$). The bath Hamiltonian is $H_B=\sum_{j}\sum_k \w_{k,j}b_{k,j}^{\dagger}b_{k,j}$, with 
frequencies $\w_{k,j}$. 
Eq.~({\ref{TotalHamiltonian}}) 
generates excitation dynamics 
within the three decoupled subspaces, spanned by $\{\ket{GG},\{\ket{XG},\ket{GX}\},\ket{XX}\}$. We are interested 
here in the single-excitation subspace (spanned by $\{\ket{XG},\ket{GX}\}$) and therefore set $\ket{XG}\equiv\ket{1}$ and 
$\ket{GX}\equiv\ket{2}$, and write the Hamiltonian governing dynamics within this subspace as
\begin{align}
H_{\mathrm{SUB}}=&{\frac{\epsilon}{2}}\sigma_z+V\sigma_x+H_B\nonumber\\
&+\sum_{j=1,2}\ketbra{j}{j}\sum_k g_{k , j}(b_{k, j}^{\dagger}+b_{k, j}).
\label{HS}
\end{align}
Here, $\epsilon=\epsilon_1-\epsilon_2$ is the energetic difference between the two sites, the Pauli operators are defined in the basis $\sigma_z=\ketbra{1}{1}-\ketbra{2}{2}$, and a term proportional to the identity has been neglected.

A standard way to proceed from this point might be to treat the system-bath interaction 
term in Eq.~({\ref{HS}}) as a perturbation, resulting 
in a master equation of Lindblad or Redfield type.~\cite{b+p} Alternatively, a polaron transformation could first be applied to 
Eq.~({\ref{HS}}), allowing one to identify a new interaction term, and resulting 
in a master equation of the form explored in Refs.~\onlinecite{mccutcheon11,yang10,nazir09, jang09, jang08}. Here, we instead employ a method originally developed by Silbey and Harris and apply a variational transformation to $H_{\mathrm{SUB}}$.~\cite{silbey84,harris85} As in the polaron case, the transformation displaces those modes in the bath coupled to the site possessing the excitation. However, unlike the full polaron transformation, we allow freedom within the variational transformation to attempt to optimise these displacements \emph{for each mode}. This will be achieved through free-energy minimisation arguments.~\cite{yarkony76,yarkony77,silbey84,harris85}

The transformed 
Hamiltonian is defined by $H_T=\e^{G}H_{\mathrm{SUB}} \e^{-G}$, with 
\beq
G=\sum_{j}\ketbra{j}{j}\sum_{k} \frac{f_{k,j}}{\w_{k,j}}(b_{k,j}^{\dagger}-b_{k,j}).
\eeq
This results in $H_T=H_0+H_I$, with free Hamiltonian
\beq
H_0=\textstyle{{\frac{1}{2}}}(\epsilon+R_1-R_2)\sigma_z+V_R \sigma_x+H_B+\textstyle{\frac{1}{2}}(R_1+R_2)\openone,
\label{H0}
\eeq
where $R_j=\sum_k f_{k,j}\w_{k,j}^{-1}(f_{k,j}-2g_{k,j})$, and the renormalised electronic coupling strength $V_R$ will be defined below. The interaction Hamiltonian now contains two 
terms, $H_I=H_{\mathrm{LINEAR}}+H_{\mathrm{DISPLACED}}$, 
with the first given given by 
\beq
H_{\mathrm{LINEAR}}=\sum_j \ketbra{j}{j}\sum_k (g_{k,j}-f_{k,j})(b_{k,j}^{\dagger}+b_{k,j}),
\eeq
which has the same form as the perturbative term in Redfield theory, but with the coupling to each mode now reduced in strength. Notice that if we were to set $f_{k,j}=g_{k,j}$ for all modes, as in the full polaron transformation, $H_{\mathrm{LINEAR}}$ would vanish as expected, though in general this is not the case. The second term in $H_I$ has the form of the perturbation used in deriving a polaron master equation, and is given by
\beq\label{Hpol}
H_{\mathrm{DISPLACED}}=V(\sigma_x B_x +\sigma_y B_y),
\eeq
written in terms of the Hermitian combinations $B_x=(1/2)(B_+ +B_- -2B)$ and 
$B_y=(i/2)(B_+-B_-)$ of the bath displacement operators operators $B_{\pm}=B_{\pm,1}B_{\mp,2}$, with
\beq
B_{\pm,j}=\exp\Big[\pm\sum_k \frac{f_{k,j}}{\w_{k,j}}(b_{k,j}^{\dagger}-b_{k,j})\Big].
\label{Bops}
\eeq
Again, it is important to note that only in the particular case of $f_{k,j}=g_{k,j}$ (for all modes) does Eq.~(\ref{Hpol}) become identical to the polaron form.

Going back to the free Hamiltonian, we see that the term generating the energy transfer in Eq.~({\ref{H0}}) now appears with a renormalised strength; $V_R=V B$, with $B=\mathrm{Tr}(B_{\pm} \rho_B)$ being the expectation value of $B_{\pm}$ with respect to the bath state. Taking a thermal equilibrium state $\rho_B=\e^{-\beta H_B}/(\mathrm{Tr}\,\e^{-\beta H_B})$ at inverse temperature $\beta=1/k_B T$ gives 
explicitly
\beq
B=\exp\Big[-\frac{1}{2}\sum_j\sum_k \frac{f_{k,j}^2}{\w_{k,j}^2}\coth\Big({\frac{\beta\,\w_{k,j}}{2}}\Big)\Big].
\eeq
If we now assume that the baths coupled to the donor and acceptor are identical, this allows us to set
 $g_{k,1}=g_{k,2}=g_{k}$, $\w_{k,1}=\w_{k,2}=\w_k$, $f_{k,1}=f_{k,2}=f_k$, and $R_1=R_2=R=\sum_k f_k \w_k^{-1}(f_k-2g_k)$. The free Hamiltonian now takes on the simpler 
form
\beq
H_0=\frac{\epsilon}{2}\sigma_z+V_R \sigma_x+R\openone+H_B,
\eeq
while the renormalisation factor becomes
\beq
B=\exp\Big[-\sum_k \frac{f_{k}^2}{\w_{k}^2}\coth(\beta\w_{k}/2)\Big].
\label{BSummation}
\eeq

\subsection{Free energy minimisation}

Currently, the parameters $\{f_{k}\}$ appearing in our transformed Hamiltonian are undetermined. We note that setting $f_{k}=0$ corresponds to 
having performed no transformation on the Hamiltonian, and as such $H_{\mathrm{DISPLACED}}=0$ while $H_{\mathrm{LINEAR}}$ remains finite, and 
the Hamiltonian takes on its original spin-boson form.~\cite{leggett87} On the other hand, as remarked earlier, by setting $f_{k}=g_k$ 
for all $k$ one finds that $H_{\mathrm{LINEAR}}=0$ while $H_{\mathrm{DISPLACED}}$ remains finite. In this case, the  
Hamiltonian corresponds to that of  
a polaron transformed spin-boson model.~\cite{wurger98,mccutcheon11, nazir09, jang09, jang08} Our aim in this work is to attempt to derive a second-order (time-local) master equation valid over as large a 
range of parameter space as possible. We achieve this by trying to find 
an interaction Hamiltonian which remains small over the greatest range of parameters. 
Therefore, rather than setting $f_{k}=0$ or $f_{k}=g_k$, we instead determine the set $\{f_{k}\}$ by free energy minimisation arguments. 

To proceed, we compute the Feynman-Bogoliuobov upper bound on the free energy,~\cite{silbey84,harris85} given by
\beq
A_B=-\frac{1}{\beta}\ln\bigl(\mathrm{Tr}\{\e^{-\beta H_{0}}\}\bigr)+\langle H_{I} \rangle_{H_{0}}+\mathcal{O}( \langle H_{I}^2 \rangle_{H_{0}}),
\label{AB}
\eeq
and related to the true free energy, $A$, via the inequality $A_B\geq A$.~\cite{fisher} By construction, the second term appearing in Eq.~({\ref{AB}}) is 
equal to zero. The variational transformation does not affect the value of the free energy of the complete system; in order to minimise the contribution from $H_I$, we therefore neglect the third term in Eq.~({\ref{AB}}) and minimise (or maximise in magnitude) the first term with respect 
to the variational parameters. We find 
\beq
A_B=R-\frac{1}{\beta}\ln\big[2 \cosh\big(\beta\eta/2)\big],
\eeq
with $\eta=\sqrt{\epsilon^2+4 V_R^2}$, and minimisation with respect to $f_k$ gives $f_k=g_k F(\w_k)$, with
\beq
F(\w_k)=\Big[1+\frac{2 V_R^2}{\eta\w_k}\tanh\big(\beta\eta/2\big)\coth\big(\beta\omega_k/2\big)\Big]^{-1}.
\label{Minimisation}
\eeq
Thus, in general, the interaction Hamiltonian contains contributions from both $H_{\mathrm{LINEAR}}$ and 
$H_{\mathrm{DISPLACED}}$. As we shall see below, in deriving a master equation utilising $H_I$ as a perturbation, we 
therefore have terms arising from both $H_{\mathrm{LINEAR}}$ and $H_{\mathrm{DISPLACED}}$, as well as from their product. 

Introducing the spectral density (assumed to be the same at each site) as $J(\w)=\sum_k |g_k|^2\delta(\w-\w_k)$ allows Eq.~({\ref{BSummation}}) to be written in integral form as
\beq
B=\exp\Big[-\int_0^{\infty}{\rm d}\w \frac{J(\w)}{\w^2}F(\w)^2\coth(\beta\w/2)\Big].
\label{RenormalisationFactor}
\eeq
Since $V_R=V B$, and $B$ is itself a function of $V_R$, the renormalised coupling strength must be solved for self-consistently.

For a sufficiently large bath of oscillators the spectral density is typically taken to be a smooth function of $\w$, with 
polynomial-like behaviour in the small $\w$ limit; $J(\w)\sim\w^s$ as $\w\rightarrow 0$. Those spectral densities with 
$s<1$, $s=1$, and $s>1$ are referred to as sub-Ohmic, Ohmic, and super-Ohmic, respectively.~\cite{leggett87} Of particular significance is the value of the renormalised interaction strength, $V_R$, found for an Ohmic environment. For the full polaron transformation, the integral in Eq.~({\ref{RenormalisationFactor}}) suffers from a well-known infra-red divergence,~\cite{leggett87,weissbook} which can be seen here by setting $F(\w)=1$ and taking $J(\w)\sim\w$. In this case, $V_R\rightarrow0$ in the polaron transformation regardless of the temperature or the strength of the system-bath coupling.
For an Ohmic environment, we are therefore forced to conclude that only \emph{incoherent} dynamics can be captured by the full polaron transformation when it is used in conjunction with a time-local master equation approach. Importantly, this is not the case for the variational theory as the function $F(\w)$ need not be equal to $1$, and hence 
both coherent and incoherent dynamics can be captured.

\section{Master equation}
\label{MasterEquation}

Having performed the variational transformation on our Hamiltonian to identify the perturbation term $H_I$, we now 
employ the time-convolutionless projection operator method~\cite{b+p} to derive a master equation governing the donor-acceptor energy transfer dynamics under the influence of the surrounding environment. 
This technique utilises a projection operator, $\mathcal{P}$, defined to satisfy 
$\mathcal{P}\chi\equiv\mathrm{Tr}_B(\chi)\otimes\rho_{\mathrm{R}}=\rho\otimes\rho_{\mathrm{R}}$, where $\chi$ is the 
density operator of the combined system-plus-environment state, $\rho$ is the reduced density operator of the system (the donor-acceptor pair in our case), and $\rho_{\mathrm{R}}$ is a reference state of the environment, which can in principle be chosen arbitrarily. Using the projection operator, 
an exact time-local master equation for $\mathcal{P}\chi$ can be derived, which involves $\mathcal{P}\chi$ itself and, in general, the complementary projection, $\mathcal{Q}\chi(0)\equiv(I-\mathcal{P})\chi(0)$, of the initial state $\chi(0)$. Importantly, in the present case, since our Hamiltonian has been transformed 
into the variational frame, we obtain a master equation involving the variationally-transformed density operator, 
$\chi_T=\e^{G}\chi\e^{-G}$. Truncating the exact expression to second order in $H_I$, 
our interaction picture master equation reads
\beq
\der{\tilde{\rho}_T}{t}= \mathrm{Tr}_B[\tilde{\mathcal{K}}(t) \mathcal{P}\tilde{\chi}_T (t)]+\mathrm{Tr}_B[\tilde{\mathcal{I}}(t)\mathcal{Q}\chi_T(0)],
\label{TotalME}
\eeq
where the homogeneous contribution is given by
\beq
\mathrm{Tr}_B[\tilde{\mathcal{K}}(t) \mathcal{P}\tilde{\chi}_T (t)]=-\int_0^{t}\mathrm{d}s\mathrm{Tr}_B[\tilde{H}_I(t),[\tilde{H}_I(s),\tilde{\rho}_T(t)\rho_{\mathrm{R}}]],
\label{HTerms}
\eeq
while the initial-state-dependent inhomogeneous term is 
\begin{align}
\mathrm{Tr}_B[\tilde{\mathcal{I}}(t)\mathcal{Q}\chi_T(0)]&=-i\mathrm{Tr}_B [\tilde{H}_I(t),\mathcal{Q}\chi_T(0)]\nonumber\\
&-\int_0^t \mathrm{d}s \mathrm{Tr}_B [\tilde{H}_I(t),[\tilde{H}_I(s),\mathcal{Q}\chi_T(0)]],
\label{IHTerms}
\end{align}
with tildes indicating operators in the interaction picture, 
$\tilde{H}_I(t)=\exp[i H_0 t]H_I \exp[-i H_0 t]$, and $\rho_T(t)=\mathrm{Tr}_B(\chi_T(t))$ being the reduced density 
operator in the variationally-transformed frame. Note that in deriving Eq.~({\ref{TotalME}}) we have utilised the fact that 
$\mathrm{Tr}_B (\tilde{H}_I(t)\rho_{\mathrm{R}})=0$.

Derivation of the final form of the master equation now 
proceeds in the usual way, and is somewhat lengthy given the various terms present in $H_I$. We therefore leave the details for Appendix~{\ref{MEDerivation}} and mention only the 
salient points here. From Eqs.~({\ref{HTerms}}) and ({\ref{IHTerms}}) it is evident that the master equation will contain two-time correlation functions arising from the product of $\tilde{H}_I(t)=\tilde{H}_{\mathrm{LINEAR}}(t)+\tilde{H}_{\mathrm{DISPLACED}}(t)$. We write $H_I=\sum_{i=1}^{4} A_i\otimes B_i$, with system operators 
$A_1=\ketbra{1}{1}$, $A_2=\ketbra{2}{2}$, $A_3=\sigma_x$, and $A_4=\sigma_y$, and bath operators
$B_i=\sum_k (g_k-f_k)(b_{k,i}^{\dagger}+b_{k,i})$ for $i=1,2$, while $B_3=V B_x$ and $B_4=V B_y$. The correlation functions appearing in the master equation can then be written as 
\begin{equation}
\Lambda_{ij}(\tau)=\mathrm{Tr}_B(\tilde{B}_i(\tau)\tilde{B}_j(0)\rho_{\mathrm{R}}), 
\end{equation}
where we choose 
a thermal equilibrium bath reference state, $\rho_{\mathrm{R}}=\e^{-\beta H_B}/(\mathrm{Tr}\,\e^{-\beta H_B})$. 

For $i=1,2$ 
these correlation functions have a similar form to those found in Redfield theory: 
\begin{align}
\Lambda_{11}(\tau)=\Lambda_{22}(\tau)=&\int_0^{\infty}\mathrm{d}\w J(\w)(1-F(\w))^2\nonumber\\
&\times\left(\cos(\w \tau)\coth(\beta\w/2)-i \sin(\w \tau)\right),
\label{WeakCorrelationFunction}
\end{align}
while $\Lambda_{12}(\tau)=\Lambda_{21}(\tau)=0$. Note that in the polaron limit ($F(\w)\rightarrow1$) these correlation functions vanish. 
For $i=3,4$, we obtain correlation functions similar in form to those of a polaron master equation:~\cite{jang08,jang09,nazir09,mccutcheon11,yang10,kolli11}
\begin{eqnarray}
\Lambda_{33}(\tau)&{}={}&(V_R^2/2)(\e^{\phi(\tau)}+\e^{-\phi(\tau)}-2),\\ 
\Lambda_{44}(\tau)&{}={}&(V_R^2/2)(\e^{\phi(\tau)}-\e^{-\phi(\tau)}), 
\end{eqnarray}
with 
\begin{align}
\phi(\tau)=&2\int_0^{\infty}\mathrm{d}\w\frac{J(\w)}{\w^2}F(\w)^2\nonumber\\
&\times\left(\cos(\w \tau)\coth(\beta\w/2)-i \sin(\w \tau)\right),
\label{PolaronCorrelationFunction}
\end{align}
while $\Lambda_{34}(\tau)=\Lambda_{43}(\tau)=0$. In the weak-coupling (Redfield) limit $F(\w)\rightarrow0$, and $\Lambda_{33}(\tau)$ and $\Lambda_{44}(\tau)$ then disappear. The 
last type of correlation function is unique to the variational theory and arises from the product of $H_{\mathrm{LINEAR}}$ and 
$H_{\mathrm{DISPLACED}}$. We have $\Lambda_{14}(\tau)=\Lambda_{42}(\tau)=i\Lambda_C(\tau)$, and 
$\Lambda_{24}(\tau)=\Lambda_{41}(\tau)=-i\Lambda_C(\tau)$, where 
\begin{align}
\Lambda_{C}(\tau)=V_R&\int_0^{\infty}\mathrm{d}\w\frac{J(\w)}{\w}F(\w)(1-F(\w))\nonumber\\
&\times\left(\sin(\w\tau)\coth(\beta\w/2)+i\cos(\w\tau)\right),
\label{CrossCorrelationFunction}
\end{align}
and $\Lambda_{31}(\tau)=\Lambda_{13}(\tau)=\Lambda_{32}(\tau)=\Lambda_{23}(\tau)=0$. These 
correlation functions are most important in the intermediate regime, where neither the 
full polaron displacement ($F(\w)=1$) nor zero displacement ($F(\w)=0$) are appropriate.

The inhomogeneous term in Eq.~(\ref{TotalME}) depends on the quantity $\mathcal{Q}\chi_T(0)$. Assuming a separable initial 
state corresponding to an excitation on the donor, $\rho(0)=\ketbra{1}{1}$, with the environment in the state 
$\rho_B(0)$, we find 
\begin{align}
\mathcal{Q}\chi_T(0)&=(1-\mathcal{P})(\e^G\ketbra{1}{1}\otimes\rho_B(0)\e^{-G})\nonumber\\
&=\ketbra{1}{1}\otimes(\rho_{\mathrm{TB}}(0)-\rho_{\mathrm{R}}),
\label{Qchi0}
\end{align}
where $\rho_{\mathrm{TB}}(0)=B_{+,1}\rho_{B}(0)B_{-,1}$ is the variationally-transformed bath initial state. Thus, we see that the inhomogeneous term vanishes if we assume our initial environmental state to be displaced such that $\rho_B(0)=B_{-,1}\rho_{\mathrm{R}}B_{+,1}$, since then $\rho_{\mathrm{TB}}(0)=\rho_{\mathrm{R}}$. We might, however, wish to consider a case where the excitation enters the system on a short enough timescale such that the bath does not have time to displace in response to it. The correct initial state would then be $\rho_B(0)=\rho_{\mathrm{R}}$, such that $\rho_{\mathrm{TB}}(0)=B_{+,1}\rho_{\mathrm{R}}B_{-,1}$. We then find 
$\mathcal{Q}\chi_T(0)\neq 0$, and so the inhomogeneous term remains. 
Details of the first and second-order inhomogeneous contributions to the master equation in this case are presented in Appendix~{\ref{MEDerivation}}.

\section{Energy Transfer Dynamics}
\label{dynamics}

Let us now use the variational master equation we have derived to investigate the dynamics of 
an excitation in the donor-acceptor system. To do so, we calculate the population in state $\ket{1}$ (the donor) as 
a function of time, $\rho_{11}(t)=(1/2)(1+\alpha_z(t))$, where $\alpha_z=\mathrm{Tr}(\sigma_z \chi(t))$. Generally, care 
must be taken when calculating expectation values from 
a master equation derived in a transformed frame. However, since $\e^G \sigma_z\e^{-G}=\sigma_z$, we find 
$\alpha_z(t)=\mathrm{Tr}(\sigma_z \chi(t))=\mathrm{Tr}(\sigma_z\e^{-G}\chi_T(t)\e^{G})=\mathrm{Tr}(\sigma_z\chi_T(t))$. Thus, the population dynamics in which we are interested is unaffected by transformation into the variational representation, and we may use our master equation for $\rho_T(t)$ directly to compute $\rho_{11}(t)$.

In the following sections we shall compare the dynamics generated by the present variational theory to that found using both 
the Redfield and polaron approaches. Redfield theory corresponds to second-order perturbation in the original system-bath interaction term, before any transformation has been applied to the Hamiltonian. Within the present formalism it can therefore be recovered by setting $F(\w)=0$. Typically, it is further assumed in this limit that the bath correlation functions 
decay on a timescale much faster than that of the excitation dynamics, which allows the integration limits in Eq.~({\ref{HTerms}}) (and also in Eq.~({\ref{ResponseFunctions}})) to be taken to infinity. Dynamics referred to as Redfield in the subsequent sections is calculated in this way. On the other hand, a polaron master equation is obtained through performing the full transformation on the original Hamiltonian. In our formalism, this corresponds to setting $F(\w)=1$. Making no further assumptions, our theory then reduces to that presented in Refs.~{\onlinecite{jang08,jang09}}. Dynamics referred to as polaron in the following is calculated in this manner.

\subsection{Super-Ohmic environments}

\begin{figure}
\begin{center}
\includegraphics[width=0.48\textwidth]{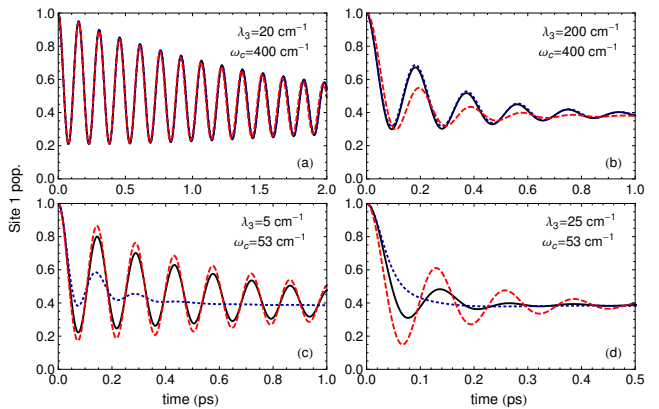}
\caption{Excitation dynamics for various reorganisation energies and cut-off frequencies, and for coupling to a super-Ohmic environment. 
Calculations using the variational master equation (black solid curves), Redfield theory (red dashed curves), and 
polaron theory (blue dotted curves) are shown. Parameters: $\epsilon=100~\mathrm{cm}^{-1}$, $V=100~\mathrm{cm}^{-1}$ and $T=300$~K.}
\label{SupOhmicPlots}
\end{center}
\end{figure}

To begin our analysis we shall first consider a donor-acceptor pair that is coupled to a super-Ohmic environment, as has been studied previously using the polaron formalism in Refs.~\onlinecite{jang08,jang09,yang10,nazir09,mccutcheon11,kolli11}. This immediately allows us to investigate how the present variational master equation
theory compares to both the Redfield and polaron forms, 
without having to worry about complications such as the polaron theory infra-red divergence, which would be an issue in the Ohmic case (see Section~\ref{OhmicEnvironments}). We take a spectral density of the form
\beq
J_3(\w)=\alpha_3 \w^3 \e^{-\w/\w_c},
\label{J3}
\eeq
where $\alpha_3$ captures the strength of the system-bath coupling and $\w_c$ is a 
phenomenological cut-off frequency. A spectral density of this form is typical in the solid-state, for example when describing coupling to acoustic phonons, see e.g. Refs.~\onlinecite{krummheuer02,ramsay10,ramsay10_2}. We also introduce the reorganisation energy, defined as 
\beq
\lambda_s=\int_0^{\infty} J_s(\w)\w^{-1}\mathrm{d}\w,
\label{ReorganisationEnergy}
\eeq
which constitutes a measure of the system-bath interaction that also accounts for the range of frequencies over which 
the bath can influence the system. For the super-Ohmic spectral density given above we find $\lambda_3=2 \alpha_3 \w_c^3$.

In Fig.~{\ref{SupOhmicPlots}} we plot the excitation population dynamics using the variational theory (solid black curves), Redfield theory (red dashed curves), 
and the polaron theory (blue dotted curves), for various values of the reorganisation energy, $\lambda_3$, and cut-off frequency, $\w_c$. In the variational 
and polaron cases we have assumed that the environment is initially in a displaced thermal equilibrium state, resulting 
in the absence of inhomogeneous terms, as discussed in Section~{\ref{MasterEquation}}. 
The effect of assuming a non-displaced initial bath 
state is discussed in Section~{\ref{WithIHTerms}}. 
In plot~(a) the reorganisation energy has a relatively small value of $\lambda_3=20~\mathrm{cm}^{-1}$, and the cut-off frequency is such that $V/\w_c<1$. For these parameters 
we expect both the Redfield and polaron theories to be valid,~\cite{wurger98,mccutcheon10_2} and hence they yield almost identical results. We also see that the variational 
theory predicts essentially the same dynamics in this relatively undemanding weak-coupling regime.

Plot (b) corresponds to the case where 
the reorganisation energy has been increased by a factor of $10$. We would not expect Redfield theory to be justified in this regime, owing to the large value of $\lambda_3$,  
and indeed we see that it predicts a strong damping of 
the system dynamics, at odds with the other approaches. 
In fact, in this regime the variational minimisation condition corresponds approximately to 
performing the full polaron displacement, i.e. $F(\w_k)\approx 1$, and so the polaron and variational theories give very similar results. This is to be expected (for a super-Ohmic bath), as for large $\lambda_3$ and small $V/\omega_c$ the full polaron transformation provides a much smaller perturbative term than the original system-bath coupling,~\cite{mccutcheon10_2} and is therefore a superior representation in which to approximate the system dynamics. 

For the parameters of plot (c), on the other hand, we now have $V/\w_c>1$, and the full polaron displacement is no longer appropriate. We therefore see that the polaron theory incorrectly predicts a strong damping of the system dynamics, even though the reorganisation energy is small. In fact, the variational minimisation condition corresponds here to performing only a weak transformation on the Hamiltonian, i.e. $F(\w_k)\approx 0$, and the variational and Redfield theories thus predict similar results.

Plot~(d) corresponds to none of the limiting 
cases discussed above, and consequently all three theories predict different dynamics. Here, we expect the variational approach to most accurately describe the true dynamics, since the minimisation condition allows for a mode-specific optimisation of the displacement of the bath, which the other theories do not. We note also that 
a recent comparison of the three theories considered here with a numerically-exact path integral approach confirmed the superiority of the variational method over Redfield and polaron for super-Ohmic environments, albeit 
for a different model system.~\cite{mccutcheon11_2}

\subsection{Ohmic environments}
\label{OhmicEnvironments}

\begin{figure}
\begin{center}
\includegraphics[width=0.45\textwidth]{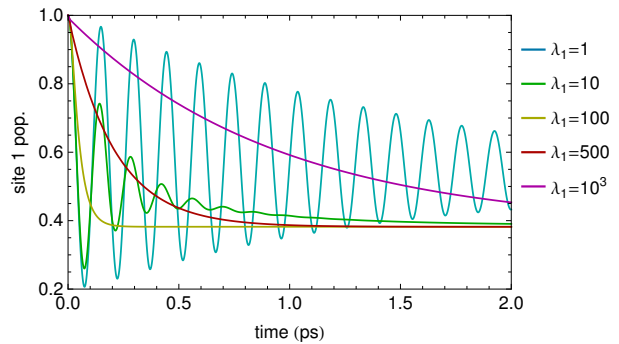}
\caption{Excitation dynamics calculated using the variational theory for coupling to an Ohmic environment, with reorganisation energies as indicated (in units of cm$^{-1}$). 
Parameters: $\epsilon=100~\mathrm{cm}^{-1}$, $V=100~\mathrm{cm}^{-1}$, $\omega_c=53~\mathrm{cm}^{-1}$, and $T=300$~K.}
\label{OhmicVariational}
\end{center}
\end{figure}

We now consider the excitation dynamics when the donor-acceptor pair are coupled to an Ohmic bath. Specifically, we take the 
overdamped Brownian oscillator form
\beq
J_1(\w)=\frac{2}{\pi}\frac{\alpha_1\, \w_c \,\w}{\w^2+\w_c^2},
\label{J1}
\eeq
which behaves linearly in the limit $\w\rightarrow 0$. From Eq.~({\ref{ReorganisationEnergy}}) we find $\lambda_1=\alpha_1$, 
which is equivalent to the reorganisation energy used in Ref.~\onlinecite{ishizaki10}.

In Fig.~{\ref{OhmicVariational}} we plot the excitation dynamics calculated from the variational theory for several values of the coupling $\lambda_1$. As in the super-Ohmic case, we assume a displaced initial bath state resulting in the absence of inhomogeneous terms. As we might expect, for small $\lambda_1$ the dynamics shows pronounced oscillations with a decaying envelope, and as $\lambda_1$ is increased, the oscillations become 
more strongly damped. Once $\lambda_1=100~\mathrm{cm}^{-1}$, however, 
the dynamics becomes entirely incoherent, and the steady state is 
reached on a timescale $\sim 0.1~\mathrm{ps}$. Interestingly, as the coupling strength is increased further, the excitation 
begins to take longer to relax towards the steady state, resulting in a decrease in the rate of excitation transfer between the two sites. This behaviour is consistent with results obtained using the hierarchal equations of motion technique.~\cite{ishizaki09b,ishizaki10}

\begin{figure}
\begin{center}
\includegraphics[width=0.45\textwidth]{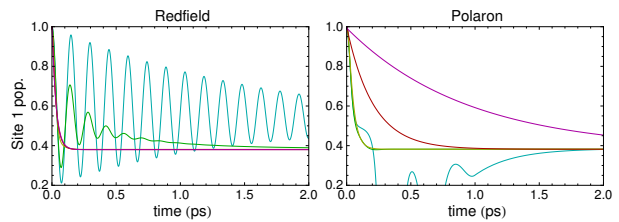}
\caption{Redfield and polaron dynamics for coupling to an Ohmic environment. Each curve corresponds to a
different reorganisation energy, as indicated in Fig.~{\ref{OhmicVariational}}. 
Parameters: $\epsilon=100~\mathrm{cm}^{-1}$, $V=100~\mathrm{cm}^{-1}$, $\omega_c=53~\mathrm{cm}^{-1}$, and $T=300$~K.}
\label{OhmicPolaronAndRedfield}
\end{center}
\end{figure}

By way of comparison, in Fig.~{\ref{OhmicPolaronAndRedfield}} we plot excitation dynamics for parameters identical to those in Fig.~{\ref{OhmicVariational}}, though now calculated using Redfield and polaron theories. As in the super-Ohmic case, we see that for small reorganisation energies, the Redfield and variational theories predict 
very similar dynamics. However, as the system-bath coupling strength becomes large, Redfield theory cannot capture the expected reduction of the transfer rate,~\cite{ishizaki09,ishizaki10} and there appears to 
be a particular coupling strength after which the steady state is always reached on a timescale $\sim 0.1~\mathrm{ps}$. 

For the Ohmic spectral density studied here, we see that polaron theory behaves quite oddly. 
As previously mentioned, a spectral density of this form presents a problem for a time-local master 
equation obtained in the polaron frame, since a complete renormalisation of the electronic transfer strength is 
predicted, $V_R\rightarrow 0$. In this situation, the polaron formalism can capture only incoherent dynamics. In fact, 
when $V_R=0$ (in either the polaron or variational theory), from Eq.~({\ref{HomoParts}}) we find a simple expression governing the time evolution of the population on site $1$: 
\beq
\der{\rho_{11}(t)}{t}=-\kappa(\epsilon,t)\rho_{11}(t)+\kappa(-\epsilon,t)(1-\rho_{11}(t)),
\eeq
where the rates determining transfer between the sites  
are given by~\cite{Note1}%~\footnote{Note that the integrand in Eq.~({\ref{kappa}}) remains finite, despite containing a factor of $B^2$, owing to a cancellation with the factor $\exp[\phi(\tau)]$ (see Eq.~({\ref{PolaronCorrelationFunction}}) with $F(\w)=1$).}
\beq
\kappa(\pm\epsilon,t)=2 V^2 \mathrm{Re}\Big[\int_0^t \mathrm{d}\tau \e^{\pm i \epsilon\tau}\left(B^2\e^{\phi(\tau)}\right)\Big].
\label{kappa}
\eeq
Thus, in the limit $V_R\rightarrow0$ we see that the polaron theory reduces to F\"{o}rster theory, though with time-dependent transfer rates,~\cite{jang02} which we expect to work well only in the strong system-environment coupling limit.~\cite{foerster59, dexter52} This is confirmed by the unphysical behaviour we see in the polaron plot for $\lambda_1=1~\mathrm{cm}^{-1}$. 
At larger coupling strengths, however, polaron theory does 
correctly predict a reduction of the transfer rate. On comparison of Figs.~{\ref{OhmicVariational}} and {\ref{OhmicPolaronAndRedfield}}, it can be seen that the 
minimisation condition [Eq.~({\ref{Minimisation}})] present in the variational theory picks out an optimised 
Hamiltonian transformation dependent upon the system-bath parameters, resulting here in 
qualitatively correct dynamical behaviour across the full range of coupling strengths.

\begin{figure}[!t]
\begin{center}
\includegraphics[width=0.48\textwidth]{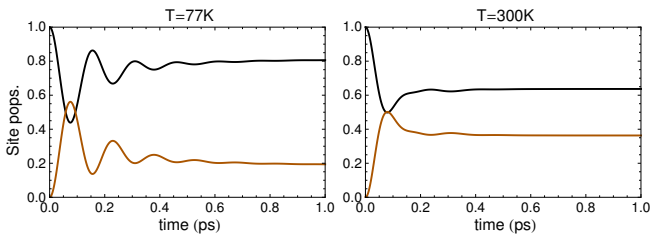}
\caption{Dynamics of the populations of BChl~1 (black) and BChl~2 (orange) for parameters corresponding to the FMO complex, at $T=77$~K (left) and $T=300$~K (right).
Parameters: $\lambda_1=35~\mathrm{cm}^{-1}$, $\w_c=106~\mathrm{cm}^{-1}$, $\epsilon=-120~\mathrm{cm}^{-1}$, $V=-87.7~\mathrm{cm}^{-1}$.}
\label{FMOPlot}
\end{center}
\end{figure}

To test the variational theory in a definite physical context, in Fig.~{\ref{FMOPlot}} we plot the energy transfer 
dynamics in a very simple model that consists of two strongly-coupled bacteriochlorophyll sites (BChl~1 and BChl~2) of the Fenna-Matthews-Olson (FMO) complex.~\cite{fenna75,adolphs06} We use our two-site Hamiltonian, Eq.~({\ref{TotalHamiltonian}}), with system parameters taken from Ref.~\onlinecite{adolphs06}  (corresponding to $\epsilon=-120~\mathrm{cm}^{-1}$ and $V=-87.7~\mathrm{cm}^{-1}$, see also Ref.~\onlinecite{ritschel11}), and consider an Ohmic spectral density [Eq.~({\ref{J1}})] with reorganisation energy $\lambda_1=35~\mathrm{cm}^{-1}$ to match Ref.~\onlinecite{ishizaki09c}. By comparison of our variational results at cryogenic temperature ($T=77$~K) and physiological temperature ($T=300$~K) with Figs.~2 and 3, respectively, in Ref.~\onlinecite{ishizaki09c}, it can be seen that the 
variational theory is in excellent agreement with the numerically-exact hierarchical approach on a timescale 
of the order $\sim 100~\mathrm{fs}$, after which time we would expect discrepancies as population begins to leak into the other  BChl sites not accounted for here.~\cite{ishizaki09c, shabani11,ritschel11,nalbach11b} 
Hence, while this may be a very simplified example, it serves to highlight the versatility of the variational 
approach, and paves the way for a more detailed study of the FMO system.

\subsection{Bath relaxation effects}
\label{WithIHTerms}

In the previous sections we assumed a displaced initial bath state when calculating dynamics using the variational and polaron theories. 
This simplified the relevant master equations, since it meant that they contained no inhomogeneous terms. This simplification relies on a separation of 
timescales in the combined system-environment dynamics, i.e. the bath relaxation must be fast compared to the energy transfer dynamics between the two sites. Thus, 
though the cut-off frequency dependence in the variational theory was seen to play an important role in the reduced dressing of the energy transfer interaction strength, the bath 
was still assumed to instantaneously adjust to its variationally displaced state. We now relax this assumption by the introduction of the inhomogeneous terms, 
the presence of which results from 
a difference between the initial bath state and that taken to be the reference state in the projection operator, and therefore (approximately) account for the influence of environmental relaxation on the state of the system. Hence, we shall now consider a non-displaced initial bath state, $\rho_B(0)=\rho_R$, 
and investigate what difference the inhomogeneous terms in our variational master equation may make, as has been studied previously in the polaron case.~\cite{jang08,jang09,kolli11} 

\begin{figure}[!t]
\begin{center}
\includegraphics[width=0.4\textwidth]{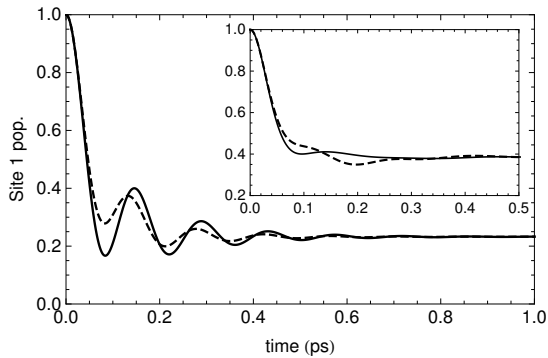}
\caption{Excitation dynamics with (dashed) and without (solid) bath relaxation terms, for a super-Ohmic spectral density. The main plot corresponds to zero temperature, while the inset shows dynamics for $T=300$~K. 
Parameters: $\lambda_3=50~\mathrm{cm}^{-1}$, $\w_c=53~\mathrm{cm}^{-1}$, $\epsilon=100~\mathrm{cm}^{-1}$, $V=100~\mathrm{cm}^{-1}$.}
\label{SupOhmicIH}
\end{center}
\end{figure}

In Fig.~{\ref{SupOhmicIH}} we plot the excitation dynamics calculated from the variational master equation for a super-Ohmic spectral density [Eq.~({\ref{J3}})] including (dashed curves) and excluding (solid curves) all first- and second-order inhomogeneous terms. The main plot is for zero temperature, while the inset corresponds to $T=300$~K. Though we find here that the effect of the inhomogeneous terms for a localised initial state is weak,~\cite{jang09}
interestingly, at zero temperature we see that the relaxation of the bath actually causes oscillations in the population dynamics to be suppressed. At $T=300$~K, the situation is somewhat different, and the dynamics is predominately incoherent without inhomogeneous terms, whereas their inclusion seems to give rise 
to small amplitude oscillations.~\cite{kolli11}

As a brief aside, at this point it is worth comparing the steady state reached at zero temperature in the variational theory to the incorrect value predicted by the non-interacting blip approximation (NIBA) for a biased system:~\cite{leggett87, weissbook} $\rho_{11}^{\mathrm{NIBA}}=(1/2)(1-\tanh(\beta\epsilon/2))$. At zero temperature, we find $\rho_{11}^{\mathrm{NIBA}}=0$, and so the NIBA predicts that all population is transferred to site~2, regardless of the reorganisation energy, or the relative values of $V$ and $\epsilon$. From Fig.~{\ref{SupOhmicIH}} we see that this is not the case in the variational theory, and we therefore expect it to better capture the corresponding dynamics in this regime.

\subsubsection*{Energy Transfer Rate}

The effect of the bath relaxation terms can also be seen in the inter-site energy transfer rates, which we explore here for an Ohmic bath [Eq.~(\ref{J1})]. To calculate these rates, we fit the time domain dynamics to the solution of the simple classical rate equation $\dot{\rho}_{11}=-\kappa_+ \rho_{11}+\kappa_-(1-\rho_{11})$, and determine the quantity $\kappa_+$, which characterises the rate at which excitation is passed from site $1$ to site $2$. Fitting quantum dynamics to 
such a rate equation is clearly dubious if there are many coherent oscillations present in the system. We shall therefore consider a small donor-acceptor interaction strength of $V=20~\mathrm{cm}^{-1}$, for which we expect the dynamics to be predominately incoherent over a large reorganisation energy range.~\cite{ishizaki10,nalbach11}

\begin{figure}[!t]
\begin{center}
\includegraphics[width=0.4\textwidth]{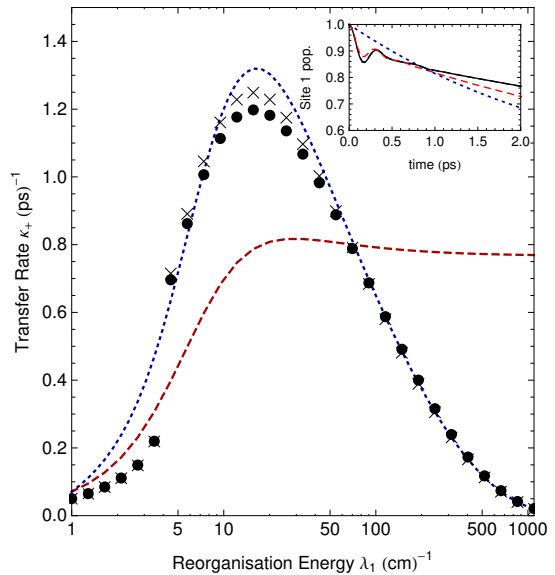}
\caption{Energy transfer rate $\kappa_+$ as a function of reorganisation energy, calculated using Redfield theory (red dashed curve), F\"{o}rster theory (blue dotted curve), and the variational theory with (black circles) and without (black crosses) 
bath relaxation terms. The inset shows the corresponding time domain dynamics for $\lambda_1=2~\mathrm{cm}^{-1}$, where the variational calculation (black solid curve) is identical both with and without inhomogeneous terms.
Parameters: $V= 20~\mathrm{cm}^{-1}$, $\epsilon=100~\mathrm{cm}^{-1}$, $\w_c=53~\mathrm{cm}^{-1}$, and $T=300$~K.}
\label{Rates}
\end{center}
\end{figure}

In the main part of Fig.~{\ref{Rates}} we plot the transfer rate, $\kappa_+$, as a function of reorganisation energy, $\lambda_1$, calculated using Redfield theory (red dashed curve), polaron or F\"{o}rster theory (blue dotted curve),~\cite{Note2} %~\footnote{In Fig.~{\ref{Rates}} we have taken the integration limit in Eq.~({\ref{kappa}}) to infinity such that the blue dotted curves correspond to the conventional Foerster theory, with time-independent rates.} 
and the variational theory with (black circles) and without (black crosses) inhomogeneous terms. The inset shows the corresponding time domain dynamics for $\lambda_1=2~\mathrm{cm}^{-1}$, confirming that the transfer is predominately incoherent, even at this small reorganisation energy. 

There are a number of interesting features to be seen in the main plot. Firstly, consistent with Fig.~{\ref{OhmicPolaronAndRedfield}}, the transfer rate calculated using Redfield theory fails to decrease significantly at large values of $\lambda_1$; as expected, Redfield theory is inadequate to describe the dynamics for large system-environment coupling strengths.~\cite{yang02,ishizaki09,ishizaki10,nalbach10_2} Secondly, and in contrast, the variational theory (both with and without inhomogeneous terms) is capable of interpolating between the small and large reorganisation energy limits, again in accord with the dynamics shown in Fig.~{\ref{OhmicPolaronAndRedfield}}. However, we note that around $\lambda_1\approx 4~\mathrm{cm}^{-1}$ the variational theory `jumps' towards the value predicted by F\"{o}rster (or polaron) theory. This behaviour can be attributed to a discontinuous change from non-zero to zero renormalised coupling $V_R$ in the variational theory, as $\lambda_1$ is increased. For the parameters of Fig.~\ref{Rates}, above $\lambda_1\approx 4~\mathrm{cm}^{-1}$ the variational minimisation condition [Eq.~(\ref{Minimisation})] suddenly reduces to $F(\omega_k)=1$, and so the polaron and variational theories then become equivalent.~\cite{Note3} %~\footnote{We note that discontinuities of this type have been observed elsewhere~\cite{harris85,zimanyiRS} when using the variational theory to describe a system coupled to an Ohmic bath} 
Finally, in the present context, perhaps the most interesting feature of Fig.~{\ref{Rates}} is that the addition of inhomogeneous terms to the variational theory serves to reduce the transfer rate for intermediate values of $\lambda_1$, bringing it closer to that predicted using the hierarchical equations of motion or path integral techniques (see Refs.~\onlinecite{ishizaki10,nalbach11} for similar plots using identical parameters).

To conclude this section we note that for most parameter regimes the inhomogeneous terms seem to be well behaved, and indeed we have just seen how they can increase the accuracy of the variational method. However, we have found that for certain parameters the inhomogeneous terms can be badly behaved, and can even cause the variational theory to predict unphysical behaviour (e.g. for very large $\lambda_1$ at $T=0$), where the behaviour is perfectly physical in their absence. We suspect that in these regimes the initial bath displacement is too great to be described accurately by only the first and second order inhomogeneous contributions to the master equation, but further work is required to fully assess this conjecture.

\section{Summary}
\label{summary}

In summary, we have presented a versatile time-local variational master equation technique, which we have used to investigate the energy transfer dynamics in a donor-acceptor pair. The master equation is constructed from a variationally-optimised interaction Hamiltonian, obtained by applying a unitary transformation to the full Hamiltonian, which results in a perturbative term that remains small over a wide range of parameter regimes. The formalism also provides a mechanism with which to (approximately) include the effects of the dynamic relaxation of the environment. One of the most important aspects of the variational master equation is its flexibility. We have seen how it naturally reduces to the well-known Redfield, polaron, and F\"{o}rster forms in the appropriate limits, and that it can also be used to explore energy transfer dynamics under conditions where those approaches are known to fail. In particular, we can capture coherent dynamics for a donor-acceptor system coupled to an Ohmic bath, something not possible in the polaron limit.

The work presented here opens up many potential avenues for future research. For example, although we have shown that the qualitative predictions of the variational master equation are generally good, it would be interesting to quantitatively assess its accuracy against known numerically-exact benchmarks.~\cite{tornow08,prior10,nalbach11,mccutcheon11_2} 
Furthermore, it may well be the case that refinements to the variational procedure, or even to the form of the unitary transformation itself, could increase the accuracy and range of validity of the method yet further. An extension to multi-site systems would allow a master equation exploration of the energy transfer dynamics valid over a range of different parameters, for example, in the full FMO complex.~\cite{fenna75,adolphs06,ritschel11} Lastly, a study utilising the formalism presented in Appendix~\ref{CorrelatedEnvironments}, which extends the theory to include spatially-correlated environmental fluctuations, could help to shed light on their influence in coherent energy transfer.

\begin{acknowledgments}

Near completion of this work we became aware of a closely related study.~\cite{zimanyiRS} We thank Eric Zimanyi and Bob Silbey for sharing a draft of their work with us, and for many interesting discussions. We are also very grateful for discussions with Alexandra Olaya-Castro, Avinash Kolli, Seogjoo Jang, and Alex Chin. This work was supported by the EPSRC and Imperial College London.

\end{acknowledgments}

\appendix

\section{Master Equation Derivation}
\label{MEDerivation}

\subsection{Homogeneous terms}

Here we present further details of the derivation of the variational master equation used in this work. 
Upon inserting the interaction Hamiltonian into Eqs.~({\ref{HTerms}}) and~({\ref{IHTerms}}), and returning to the Schr\"{o}dinger picture, 
we find a master equation of similar form to Eq.~({\ref{TotalME}}), 
$\der{\rho_T}{t}= \mathrm{Tr}_B[\mathcal{K}(t) \mathcal{P}\chi_T (t)]+\mathrm{Tr}_B[\mathcal{I}(t)\mathcal{Q}\chi_T(0)]$, 
where the homogenous contribution can be written
%
%\begin{align}
%&\mathrm{Tr}_B[\mathcal{K}(t) \mathcal{P}\chi_T (t)]=-i[H_S,\rho_T(t)]\nonumber\\
%&-{\textstyle{\frac{1}{2}}}\sum_{ij}\sum_{\w}\gamma_{ij}(\w,t)[A_i,A_j(\w)\rho_T(t)-\rho_T(t)A_j^{\dagger}(\w)]\nonumber\\
%&-i\sum_{ij}\sum_{\w}S_{ij}(\w,t)[A_i,A_j(\w)\rho_T(t)+\rho_T(t)A_j^{\dagger}(\w)],
%\label{HomoParts}
%\end{align}
\begin{align}
&\mathrm{Tr}_B[\mathcal{K}(t) \mathcal{P}\chi_T (t)]=-i[H_S,\rho_T(t)]\nonumber\\
&-{\textstyle{\frac{1}{2}}}\sum_{ij}\sum_{\w}\gamma_{ij}(\w,t)[A_i,\Delta_{j,\w}\rho_T(t)-\rho_T(t)\Delta_{j,\w}^{\dagger}]\nonumber\\
&-i\sum_{ij}\sum_{\w}S_{ij}(\w,t)[A_i,\Delta_{j,\w}\rho_T(t)+\rho_T(t)\Delta_{j,\w}^{\dagger}],
\label{HomoParts}
\end{align}
with $H_S=\textstyle{\frac{1}{2}}\epsilon\sigma_z+V_R\sigma_x+R\openone$. The indices $i$ and $j$ ($i,j=1,2,3,4$) label system operators 
$A_1=\ketbra{1}{1}$, $A_2=\ketbra{2}{2}$, $A_3=\sigma_x$, and $A_4=\sigma_y$.  
In deriving Eq.~({\ref{HomoParts}}) we have utilised the following decomposition of the system operators, 
\begin{equation}
\Delta_{i,\w}=\sum_{E'-E=\w}\ketbra{E}{E}A_i\ketbra{E'}{E'},
\label{Decomposition}
\end{equation}
where the summation runs over all pairs of eigenvalues of $H_S$ having a fixed energy difference $\w$. Using   
Eq.~({\ref{Decomposition}}) the transformation into the interaction picture is easily achieved since we have $\tilde{\Delta}_{i,\w}(t)=\exp[i H_S t]\Delta_{i,\w}\exp[-i H_S t]=\Delta_{i,\w}\exp[-i \w t]$, while  
$A_i=\sum_{\w}\Delta_{i,\w}$. Given our form for $H_S$ we find that the $\w$ summation runs over the three values 
$\w=0,\pm\eta$, where $\eta=\sqrt{\epsilon^2+4V_R^2}$ as in the main text, and the eigenoperators are given by
%The frequency summations run over $\w=0,\pm\eta$, with the system eigenoperators given by 
$\Delta_{1,\eta}=-\cos\theta\sin\theta \ketbra{-}{+}$, $\Delta_{1,0}=\sin^2\theta\ketbra{-}{-}+\cos^2\theta\ketbra{+}{+}$, 
$\Delta_{2,\eta}=\cos\theta\sin\theta \ketbra{-}{+}$, $\Delta_{2,0}=\cos^2\theta\ketbra{-}{-}+\sin^2\theta\ketbra{+}{+}$, 
$\Delta_{3,\eta}=\cos2\theta\ketbra{-}{+}$, $\Delta_{3,0}=\sin2\theta(\ketbra{+}{+}-\ketbra{-}{-})$, $\Delta_{4,\eta}=i \ketbra{-}{+}$, and 
$\Delta_{4,0}=0$. In all cases $\Delta_{i,-\eta}=\Delta_{i,\eta}^{\dagger}$, and eigenstates of $H_S$ are defined to satisfy 
$H_S\ket{\pm}=(1/2)(2R\pm\eta)\ket{\pm}$. The angle $\theta=(1/2)\arctan(2 V_R/\epsilon)$ characterises the relative 
strength of the renormalised excitonic transfer interaction to the energy mismatch between the donor and acceptor.

The time-dependent rates and energy shifts appearing in Eq.~({\ref{HomoParts}}) are given by $\gamma_{ij}(\w,t)=2\mathrm{Re}[K_{ij}(\w,t)]$ and $S_{ij}(\w,t)=\mathrm{Im}[K_{ij}(\w,t)]$, respectively, defined in terms of the response functions
\beq
K_{ij}(\w,t)=\int_0^t \Lambda_{ij}(\tau)\e^{i\w t}\mathrm{d}\tau,
\label{ResponseFunctions}
\eeq
which depend on the bath correlation functions $\Lambda_{ij}(\tau)=\mathrm{Tr}(\tilde{B}_i(\tau)\tilde{B}_j(0)\rho_{\mathrm{R}})$. Here, we label $B_i=\sum_k (g_k-f_k)(b_{k,i}^{\dagger}+b_{k,i})$ for $i=1,2$, while $B_3=V B_x$ and $B_4=V B_y$. These correlation functions are given by Eqs.~({\ref{WeakCorrelationFunction}})~-~({\ref{CrossCorrelationFunction}}) of the main text.

\subsection{Inhomogeneous terms}
\label{MEIHParts}

The inhomogeneous terms in the master equation depend on the quantity $\mathcal{Q}\chi_T(0)$. If we wish to consider the initial 
state $\chi(0)=\ketbra{1}{1}\otimes \rho_{\mathrm{R}}$, we find
\begin{align}
\mathcal{Q}\chi_T(0)&=(1-\mathcal{P})(\e^G\ketbra{1}{1}\otimes\rho_{\mathrm{R}}\e^{-G})\nonumber\\
&=\ketbra{1}{1}\otimes(\rho_{\mathrm{TR}}-\rho_{\mathrm{R}}).
\label{AppQchi0}
\end{align}
where $\rho_{\mathrm{TR}}=B_{+,1}\rho_{\mathrm{R}}B_{-,1}$ is a transformed bath reference state. 
Using Eq.~({\ref{IHTerms}}) we find that the inhomogeneous terms take the form
\begin{align}
&\mathrm{Tr}_B[\mathcal{I}(t)\mathcal{Q}\chi_T(0)]=-i\sum_i\sum_{\w} \Gamma_i (t)\e^{i\w t}[A_i, \Delta_{1,\w}]\nonumber\\
&-\sum_{ij}\sum_{\w\w'}\big(\textstyle{\frac{1}{2}}\gamma_{ij}^{(d)}(\w',t)\cos\w t-S_{ij}^{(d)}(\w',t)\sin\w t\big)\nonumber\\
&\hskip+18mm\times[A_i,\Delta_{j,\w'}\Delta_{1\w}-\Delta_{1,\w}^{\dagger}\Delta_{j,\w'}^{\dagger}]\nonumber\\
&-i\sum_{ij}\sum_{\w\w'}\big(\textstyle{\frac{1}{2}}\gamma_{ij}^{(d)}(\w',t)\sin\w t+S_{ij}^{(d)}(\w',t)\cos\w t\big)\nonumber\\
&\hskip+18mm\times[A_i,\Delta_{j,\w'}\Delta_{1,\w}+\Delta_{1,\w}^{\dagger}\Delta_{j,\w'}^{\dagger}],
\label{IHomoParts}
\end{align}
in the Schr\"{o}dinger picture, where $\Gamma_i(t)=\mathrm{Tr}_B(\tilde{B}_i(t)\rho_{\mathrm{TR}})$. In a similar way to those appearing in the homogeneous terms, 
the rates and energy shifts are defined as $\gamma_{ij}^{(d)}(\w,t)=2\mathrm{Re}[K_{ij}^{(d)}(\w,t)]$ and 
$S_{ij}^{(d)}(\w,t)=\mathrm{Im}[K_{ij}^{(d)}(\w,t)]$, this time in terms of the response functions
\beq
K_{ij}^{(d)}(\w,t)=\int_0^t \Lambda_{ij}^{(d)}(\tau,t)\e^{i\w t}\mathrm{d}\tau,
\eeq
with $\Lambda_{ij}^{(d)}(\tau,t)=\mathrm{Tr}_B(\tilde{B}_i(t)\tilde{B}_j(t-\tau)\rho_{\mathrm{TR}})-\Lambda_{ij}(\tau)$.

In order to evaluate the quantities $\Gamma_i(t)$ and $\Lambda_{ij}^{(d)}(\tau,t)$ we utilise the cyclic invariance of the trace and insertions 
of the identity operator, $I=B_{+,1}B_{-,1}$, to write
\beq
\Gamma_i(t)=\mathrm{Tr}_B(\tilde{B}_i(t)\rho_{\mathrm{TR}})=\mathrm{Tr}_B(\tilde{\mathcal{B}}_i(t)\rho_{\mathrm{R}})
\eeq 
and
\beq
\Lambda_{ij}^{(d)}(\tau,t)=\mathrm{Tr}_B(\tilde{\mathcal{B}}_i(t)\tilde{\mathcal{B}}_j(t-\tau)\rho_{\mathrm{R}})-\Lambda_{ij}(\tau),
\eeq
where $\tilde{\mathcal{B}}_i(t)=B_{-,1}\tilde{B}_i(t)B_{+,1}$. With reference to Eq.~({\ref{Bops}}), we find 
$\tilde{\mathcal{B}}_1(t)=\tilde{B}_1(t)+C(t)$, $\tilde{\mathcal{B}}_2(t)=\tilde{B}_2(t)$, 
$\tilde{\mathcal{B}}_3(t)=(V/2)(\tilde{B}_+(t) C_+(t)+\tilde{B}_-(t)C_-(t)-2B)$, and 
$\tilde{\mathcal{B}}_4(t)=(i V/2)(\tilde{B}_+(t) C_+(t)-\tilde{B}_-(t)C_-(t))$, where 
\beq
C_1(t)=2\int_0^{\infty}{\rm d}\w \frac{J(\w)}{\w}F(\w)(1-F(\w))\cos\w t
\eeq
and $C_{\pm}(t)=\exp[\pm i \psi(t)]$, with 
\beq
\psi(t)=2\int_0^{\infty}{\rm d}\w \frac{J(\w)}{\w^2}F(\w)^2\sin\w t.
\eeq
We therefore find $\Gamma_1(t)=C_1(t)$, $\Gamma_2(t)=0$, while $\Gamma_3(t)=V_R(\cos\psi(t)-1)$, 
and $\Gamma_4(t)=-V_R \sin\psi(t)$. In a similar way, we find $\Lambda_{11}^{(d)}(\tau,t)=C_1(t)C_1(t-\tau)$ 
while $\Lambda_{12}^{(d)}(\tau,t)=\Lambda_{21}^{(d)}(\tau,t)=\Lambda_{22}^{(d)}(\tau,t)=0$. Those quantities arising from 
$H_{\mathrm{DISPLACED}}$ are given by
\begin{align}
\Lambda_{33}^{(d)}(\tau,t)=\frac{V_R^2}{2}\Big(&\e^{\phi(\tau)}\big(\cos[\psi(t)-\psi(t-\tau)]-1\big)\nonumber\\
+&\e^{-\phi(\tau)}\big(\cos[\psi(t)+\psi(t-\tau)]-1\big)\nonumber\\
-&2\big(\cos\psi(t)+\cos\psi(t-\tau)-2\big)\Big),
\end{align}
\begin{align}
\Lambda_{44}^{(d)}(\tau,t)=&\frac{V_R^2}{2}\Big(\e^{\phi(\tau)}\big(\cos[\psi(t)-\psi(t-\tau)]-1\big)\nonumber\\
-&\e^{-\phi(\tau)}\big(\cos[\psi(t)+\psi(t-\tau)]-1\big)\Big),
\end{align}
\begin{align}
\Lambda_{34}^{(d)}&(\tau,t)=\frac{V_R^2}{2}\Big(\e^{\phi(\tau)}\sin[\psi(t)-\psi(t-\tau)]\nonumber\\
-&\e^{-\phi(\tau)}\sin[\psi(t)+\psi(t-\tau)]+2\sin\psi(t-\tau)\Big),
\end{align}
and
\begin{align}
\Lambda_{43}^{(d)}(\tau,t)&=\frac{V_R^2}{2}\Big(-\e^{\phi(\tau)}\sin[\psi(t)-\psi(t-\tau)]\nonumber\\
-&\e^{-\phi(\tau)}\sin[\psi(t)+\psi(t-\tau)]+2\sin\psi(t)\Big),
\end{align}
where $\phi(t)$ is defined in Eq.~(\ref{PolaronCorrelationFunction}).
Correlation functions of this type coming from the product of $H_{\mathrm{DISPLACED}}$ and 
$H_{\mathrm{LINEAR}}$ are given by 
\begin{align}
\Lambda_{13}^{(d)}(\tau,t)&=i \Lambda_C(\tau)\sin\psi(t-\tau)+\Gamma_1(t)\Gamma_3(t-\tau),\\
\Lambda_{23}^{(d)}(\tau,t)&=-i \Lambda_C(\tau)\sin\psi(t-\tau),\\
\Lambda_{31}^{(d)}(\tau,t)&=-i \Lambda_C(\tau)\sin\psi(t)+\Gamma_1(t-\tau)\Gamma_3(t),\\
\Lambda_{32}^{(d)}(\tau,t)&=i \Lambda_C(\tau)\sin\psi(t),
\end{align}
and
\begin{align}
\Lambda_{14}^{(d)}(\tau,t)&=i\Lambda_C(\tau)[\cos\psi(t-\tau)-1]+\Gamma_1(t)\Gamma_4(t-\tau),\\
\Lambda_{24}^{(d)}(\tau,t)&=-i\Lambda_C(\tau)[\cos\psi(t-\tau)-1],\\
\Lambda_{41}^{(d)}(\tau,t)&=-i\Lambda_C(\tau)[\cos\psi(t)-1]+\Gamma_1(t-\tau)\Gamma_4(t),\\
\Lambda_{42}^{(d)}(\tau,t)&=i\Lambda_C(\tau)[\cos\psi(t)-1\big].
\end{align}

\section{Correlated Environments}
\label{CorrelatedEnvironments}

Here we extend the variational formalism presented in Section~{\ref{Model}} to include the effects of 
spatial correlations in the environmental fluctuations. To do so, rather than assuming that each site is coupled to a separate bath of oscillators, we instead couple each site to a \emph{common} environment. The total Hamiltonian now reads
\begin{align}
H=&\sum_{j=1,2}\epsilon_j \ketbrai{X}{X}{j}+V\big(\ketbra{XG}{GX}+\ketbra{GX}{XG}\big)\nonumber\\
+&\sum_{j=1,2}\ketbrai{X}{X}{j}\sum_k (g_{k , j}b_{k}^{\dagger}+g_{k , j}^*b_{k})+H_{B},
\label{AppTotalHamiltonian}
\end{align}
with $H_B=\sum_k \w_k b_k^{\dagger}b_k$, and we note that we now have just one creation operator for each mode, $b_k^{\dagger}$. 
We make the spatial separation between the sites explicit with position dependent phase factors in the coupling constants. 
We write $g_{k,j}=g_k\e^{i \vec{k}\cdot\vec{r}_j}$ with $\vec{r}_j$ the position of site $j$ (this further 
assumes each site is coupled to the environment with the same strength). The Hamiltonian describing dynamics within the single excitation subspace 
has a similar form as before,
\begin{align}
H_{\mathrm{SUB}}=&\frac{\epsilon}{2}\sigma_z+V\sigma_x+H_B\nonumber\\
&+\sum_{j=1,2}\ketbra{j}{j}\sum_k (g_{k , j}b_{k}^{\dagger}+g_{k , j}^*b_{k}),
\label{AppHS}
\end{align}
with the Pauli operators defined in the same way as in the uncorrelated case.

The variationally-transformed Hamiltonian is once again defined as $H_T=\e^G H_{\mathrm{SUB}}\e^{-G}$, but now we have~\cite{mccutcheon10} 
\beq
G=\sum_{j}\ketbra{j}{j}\sum_{k} \Big(\frac{f_{k,j}}{\w_{k}}b_{k}^{\dagger}-\frac{f_{k,j}^*}{\w_{k}}b_{k}\Big),
\label{AppG}
\eeq
with the variational parameters also containing phase factors, $f_{k,j}=f_k \e^{i \vec{k}\cdot\vec{r}_j}$. The transformed 
Hamiltonian reads $H_T=H_0+H_I$, where
\beq
H_0=\textstyle{\frac{1}{2}}\epsilon\sigma_z+V_R \sigma_x+H_B+\textstyle{\frac{1}{2}}R\openone,
\label{AppH0}
\eeq
and $H_I=H_{\mathrm{LINEAR}}+H_{\mathrm{DISPLACED}}$. Here,
\beq
H_{\mathrm{LINEAR}}=\sum_j \ketbra{j}{j}\sum_k (g_{k}-f_{k})(\e^{i \vec{k}\cdot\vec{r}_j}b_{k}^{\dagger}+\e^{-i \vec{k}\cdot\vec{r}_j}b_{k}),
\eeq
and $H_{\mathrm{DISPLACED}}=V(\sigma_x B_x +\sigma_y B_y)$, with $B_x=(1/2)(B_+ +B_- -2B)$ and 
$B_y=(i/2)(B_+-B_-)$ as before, but now
\beq
B_{\pm}=\prod_k D\Big(\pm\frac{f_k}{\w_k}(\e^{i\vec{k}\cdot\vec{r}_1}-\e^{i\vec{k}\cdot\vec{r}_2})\Big),
\eeq
where $D(x_k)=\exp[x_k b_k^{\dagger}-x_k^* b_k]$ is a displacement operator. Importantly, the renormalisation factor, 
$B=\mathrm{Tr}(B_{\pm} \rho_B)$ now depends on the distance between the two sites,
\beq
B=\exp\Big[-\sum_k \frac{f_k^2}{\w_k^2}(1-\cos (\vec{k}\cdot\vec{d}))\coth(\textstyle{\frac{\beta \w_k}{2}})\Big]
\eeq
where $\vec{d}=\vec{r}_1-\vec{r}_2$. 
The free energy minimisation is performed in the same way as in Section~{\ref{Model}} and again gives
$f_k=g_kF(\w_k)$, but now we find 
\beq
F(\w_k)=\Big(1+\frac{2 V_R^2}{\eta \w_k}(1-\cos (\vec{k}\cdot\vec{d}))\tanh(\textstyle{\frac{\beta\eta}{2}})\coth(\textstyle{\frac{\beta \w_k}{2}})\Big)^{-1}.
\eeq

The derivation of the variational master equation then proceeds in the same way as in the uncorrelated case, 
and results in a form identical to Eqs.~({\ref{HomoParts}}) and ({\ref{IHomoParts}}). The difference, however, is that the functions $\Lambda_{ij}(\tau)$, $\Lambda_{ij}^{(d)}(\tau,t)$, and $\Gamma_i(t)$, now also have the potential to depend on $\vec{d}$. We find 
\begin{align}
&\Lambda_{12}(\tau)=\Lambda_{21}(\tau)=\sum_k g_k^2(1-F(\w_k))^2\nonumber\\
&\times\big(\cos(\w_k \tau-\vec{k}\cdot\vec{d})\coth(\beta\w_k/2)-i \sin(\w_k \tau-\vec{k}\cdot\vec{d})\big),
\label{AppWeakCorrelationFunction}
\end{align}
while $\Lambda_{11}(\tau)=\Lambda_{22}(\tau)$ and are found by setting $\vec{d}=0$ in 
Eq.~({\ref{AppWeakCorrelationFunction}}). Once again $\Lambda_{33}(\tau)=(V_R^2/2)(\e^{\phi(\tau)}+\e^{-\phi(\tau)}-2)$,  
$\Lambda_{44}(\tau)=(V_R^2/2)(\e^{\phi(\tau)}-\e^{-\phi(\tau)})$, and $\Lambda_{34}(\tau)=\Lambda_{43}(\tau)=0$, but now we have
\begin{align}
\phi(\tau)=& 2\sum_k\frac{g_k^2}{\w_k^2}F(\w_k)^2(1-\cos(\vec{k}\cdot\vec{d}))\nonumber\\
&\times\big(\cos(\w_k \tau)\coth(\beta\w_k/2)-i \sin(\w_k \tau)\big).
\label{AppPolaronCorrelationFunction}
\end{align}
The final correlation functions appearing in the homogeneous terms are given by $\Lambda_{14}(\tau)=\Lambda_{42}(\tau)=i\Lambda_C(\tau)$ and 
$\Lambda_{24}(\tau)=\Lambda_{41}(\tau)=-i\Lambda_C(\tau)$ where 
\begin{align}
\Lambda_{C}(\tau)=V_R&\sum_k\frac{g_k^2}{\w_k}F(\w_k)(1-F(\w_k))(1-\cos(\vec{k}\cdot\vec{d}))\nonumber\\
&\times\big(\sin(\w_k\tau)\coth(\beta\w_k/2)+i\cos(\w_k\tau)\big),
\label{AppCrossCorrelationFunction}
\end{align}
and $\Lambda_{31}(\tau)=\Lambda_{13}(\tau)=\Lambda_{32}(\tau)=\Lambda_{23}(\tau)=0$. 

To find the quantities appearing in the inhomogeneous terms, we must first find the transformed bath operators, 
$\tilde{\mathcal{B}}_i(t)=B_{-,1} \tilde{B}_i(t)B_{+,1}$, where $B_{\pm,1}=\prod_k D(f_{k,1})$. We now have 
$\tilde{\mathcal{B}}_i(t)=\tilde{B}_i(t)+C_i (t)$, for $i=1,2$, where 
\beq
C_2(t)=2\sum_k\frac{g_k^2}{\w_k}F(\w_k)(1-F(\w_k))\cos(\w_k t -\vec{k}\cdot\vec{d}),
\label{C2}
\eeq
and $C_1(t)$ is found by setting $\vec{d}=0$ in Eq.~({\ref{C2}}). Lastly, $\tilde{\mathcal{B}}_3(t)=(V/2)(\tilde{B}_+(t) C_+(t)+\tilde{B}_-(t)C_-(t)-2B)$, and 
$\tilde{\mathcal{B}}_4(t)=(i V/2)(\tilde{B}_+(t) C_+(t)-\tilde{B}_-(t)C_-(t))$, where
\beq
\psi(t)=2\sum_k\frac{g_k^2}{\w_k^2}F(\w_k)^2\big(\sin(\w_k t)+\sin(\w_k t-\vec{k}\cdot\vec{d})\big).
\eeq
With these transformed operators the functions $\Gamma_i(t)$ and $\Lambda_{ij}^{(d)}(\tau,t)$ can be found in a similar 
manner to Appendix~{\ref{MEDerivation}}.

%\bibliography{refs}

%merlin.mbs 2010-03-15 4.21a (PWD, AO, DPC)

\end{document}